%% file: make_astro.tex
\documentclass{mpe_report}

\usepackage{psfig,graphicx,epsfig}
\usepackage{color}
\usepackage{amsmath,amssymb,epic,eepic,array}

\begin{document}

\pagenumbering{arabic}
\setcounter{page}{216}

 \renewcommand{\FirstPageOfPaper }{216}\renewcommand{\LastPageOfPaper }{218}\include{./mpe_report_aulbert}          \clearpage

\end{document}

%% file: mpe_report_aulbert.tex
\title{Finding Binary Millisecond Pulsars with the Hough Transform}
\author{Carsten Aulbert}  
\institute{Albert-Einstein-Institute\\Am M\"uhlenberg~1, Potsdam, Germany}
\maketitle

\begin{abstract}
  The Hough transformation has been used successfully for more than four
decades. Originally used for tracking particle traces in bubble
chamber images, this work shows a novel approach turning the initial idea
into a powerful tool to \textbf{incoherently} detect millisecond pulsars
in binary orbits.

This poster presents the method used, a discussion on how to treat
the time domain data from radio receivers and create the input "image"
for the Hough transformation, details about the advantages and disadvantages
of this approach, and finally some results from pulsars in 47 Tucanae.
\end{abstract}

\section*{Introduction}

The problem of discovering millisecond pulsars in binary orbits is
apparent. Not only the parameter range for isolated pulsars has to be
addressed, but --- at least --- five Keplerian parameters have
to be included additionally. A simple Fourier analysis is not sufficient
for this class of pulsars, since the signal is smeared out over a
large frequency band due to Doppler's effect (ref. Figure.~\ref{fig:simpleHough}).

\begin{figure}
\centerline{\psfig{file=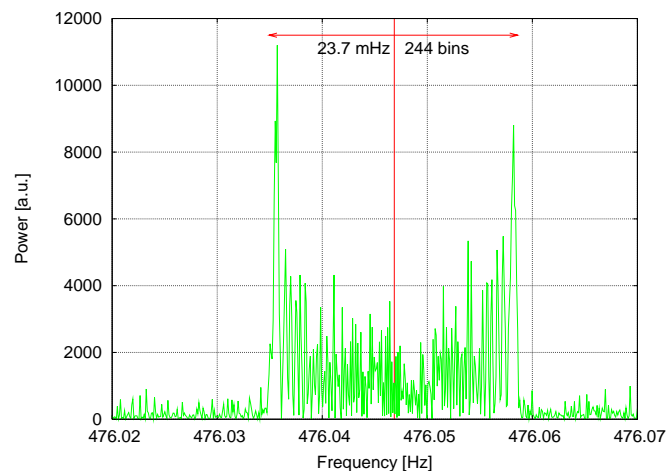,width=8.8cm,clip=} }
\caption{Power spectrum of the binary pulsar 47~Tuc~J.}
\label{fig:TucJ}
\end{figure}

Over the past decades quite a few search algorithms have been
devised to counter this problem, e.g.\ the very successful
``acceleration search'' \cite{Middleditch:1984} or the sideband search
\cite{2003ApJ...589..911R}. However, these approaches work purely on coherent data
sets, thus they are ultimately limited by the maximum possible length
of an observation

We would like to present an alternative approach to search for weak
pulsars in a binary orbit, which utilizes data taken in several
observation runs. In the following section we will briefly explain how
the Hough transformation works, and how it can be applied to search
for binary, millisecond pulsars before 
summarizing some of the results obtained for 47~Tucanae.

\section*{The Hough transformation}

Paul Hough developed and patented the algorithm around 1960. The
original algorithm was designed to search
for straight particle traces in bubble chamber images by simply
turning the usual line
equation {\boldmath $y=ax+b$} into a parameter space equation {\boldmath$
b=y-ax$}.

From these two equations it is obvious that a line in the original
space (``\textsl{xy}-space'') corresponds to a single point in the
parameter space (``\textsl{ab}-space''). On the other hand a single point in
\textsl{xy}-space is represented by a line in \textsl{ab}-space. Although sounding
trivial, it actually solves a complex problem: Identifying
parameters of lines found in (pixel) images.

\begin{figure}[h]
    \centerline{\psfig{file=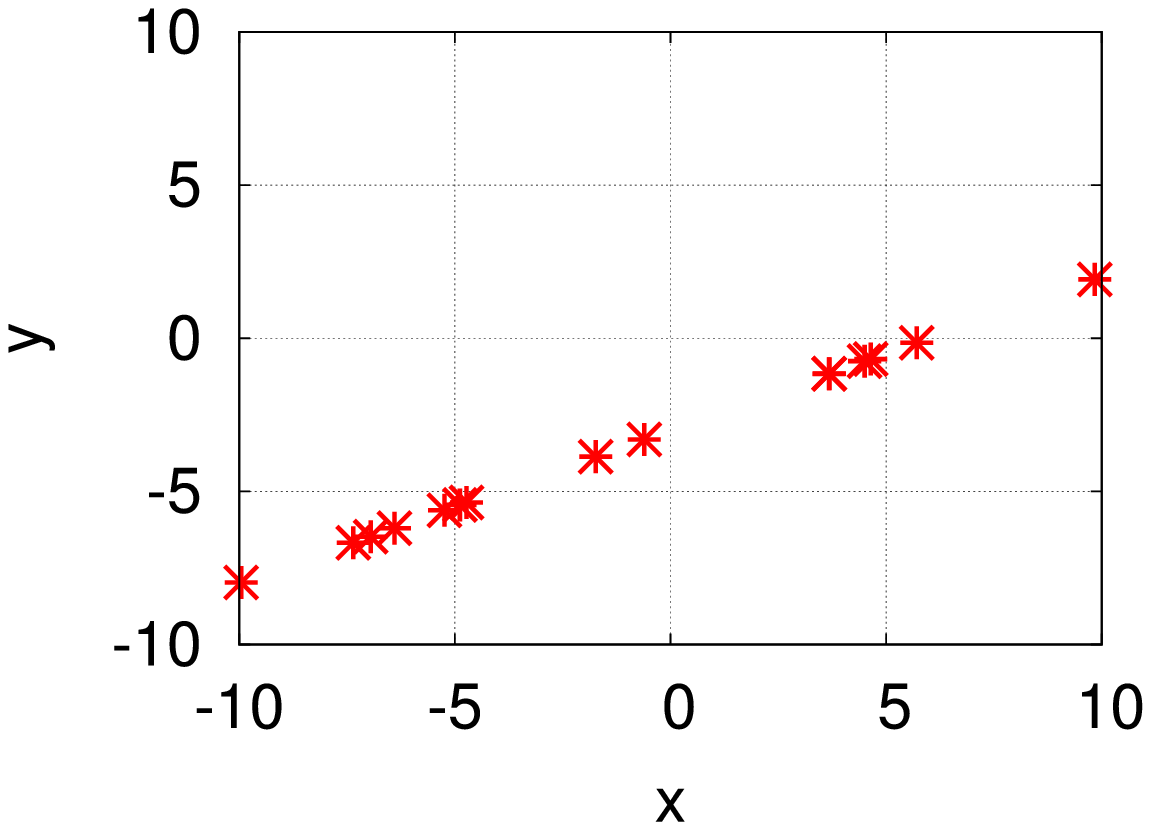,width=8.8cm,clip=} } 
    \centerline{\psfig{file=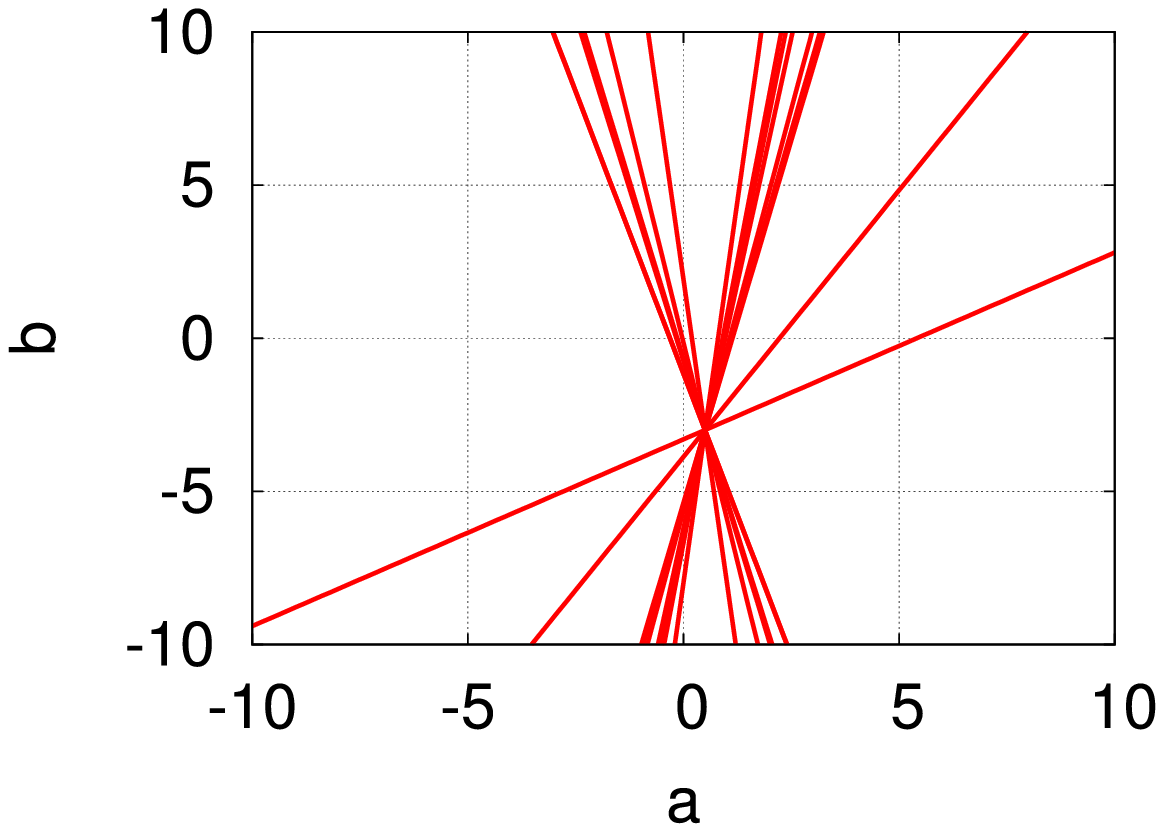,width=8.8cm,clip=} }
  \caption{Points from a line in \textsl{xy}-space transform into
     lines intersecting in a single \textsl{ab}-point.}
  \label{fig:simpleHough}
\end{figure}

Figure~\ref{fig:simpleHough} shows what happens when transforming many points
from a single line. For each of these
points a line is drawn in \textsl{ab}-space and since each of these points
belong to the same \textsl{xy}-line, their lines are required to intersect in a single
point in \textsl{ab}-space. Naturally, random points in \textsl{xy}-space convert to
random lines in \textsl{ab}-space only adding up to a random background.

\begin{figure}
  \centering
  \centerline{\psfig{file=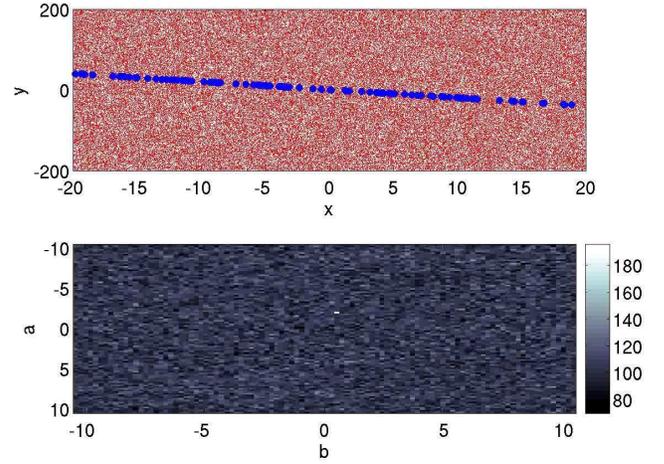,width=8.8cm,clip=} }
  \caption{Recovering a hidden line (top,blue) with a SNR of about 20
    (tiny white point (bottom)).}
  \label{fig:hiddenLine}
\end{figure}

Tracking millions of lines along with their intersections is costly in
terms of computing power and memory, thus we chose to discretize the
\textsl{ab}-space into rectangular pixels and simply count how many lines are
drawn through each pixel. This approach allows us to discover the
blue ``line'' in Figure~\ref{fig:hiddenLine}, where only 95~signal
points are hidden beneath 200,000~noise points while still be detectable at an SNR of about 20!

For binary pulsars, the frequency signature can be described by
$$f(t)=f_p(1+\lambda_o\cos(2\pi f_o-\Phi_o)),$$
where $f_p$ is the intrinsic frequency of the pulsar, $f_o$ the
orbital frequency, $\lambda_o$ the (dimensionless) frequency amplitude
due to Doppler modulation and $\Phi_o$ the orbital phase at
$t=0$. This equation can be recast into a line equation
\begin{eqnarray}
C&=&Ax+By, \mbox{ where}\nonumber\\
A&=&\sin(2\pi f_o-\Phi_o),\nonumber\\
B&=&\cos(2\pi f_o-\Phi_o),\nonumber\\
C&=&(f(t)-f_p)/f_p,\nonumber\\
x&=&f_p\lambda_o\sin\Phi_o,\nonumber\\
y&=&f_p\lambda_o\cos\Phi_o.\nonumber
\end{eqnarray}

Here, we split the variables into two types: 
\begin{description}
\item[{\boldmath $\lambda_o$}, {\boldmath $\Phi_o$}:] These variables are considered in the Hough
  transform, i.e.\ the resulting Hough planes (formerly
  \textsl{ab}-space) are polar diagrams in these parameters.
\item[{\boldmath $f_p$}, {\boldmath $f_o$}:] Called ``external'' variables,
  because they appear only as input parameters to the Hough
  transform and are varied outside the scope of it. This
  separation allows us to distribute the work onto a cluster of
  computers.
\end{description}

So far, we have not said anything about our ``input image''. From the
time domain data\footnote{Kindly made available for us by the pulsar
  group at Jodrell Bank, Manchester, UK} we use small pieces of the
data (typically a few minutes), create a power spectrum for each piece
and select peaks from the spectra. Then, we draw the registered peaks
into a time-frequency plane which serves as our input for the Hough
transformation.

\section*{47~Tucanae results}

\begin{figure}
  \centering
  \centerline{\psfig{file=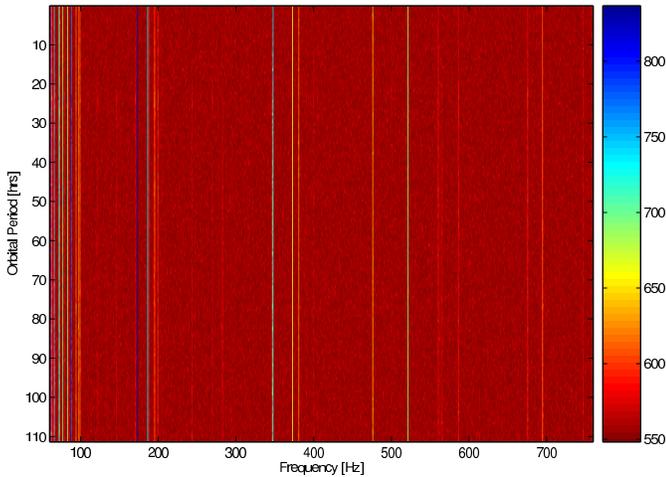,width=8.8cm,clip=} }
  \caption{47~Tucanae global result map}
  \label{fig:fullresult}
\end{figure}

In this section we summarize some results
obtained from 47~Tucanae data. Figure~\ref{fig:fullresult} shows the
outcome obtained for the full parameter space ($f_p$, $P_o=f_o^{-1}$).

The maximum number counts (of how many lines are drawn through any
pixel) found in each of the about
{\boldmath $10^9$} computed Hough planes are color coded in this
image. One can clearly see the
signatures of several noise sources (lines at and below
100~Hz as well as pulsars (173~Hz,
186~Hz) and their harmonics at twice, thrice, \dots\ of the
fundamental frequency.

\begin{figure}
  \centering
  \centerline{\psfig{file=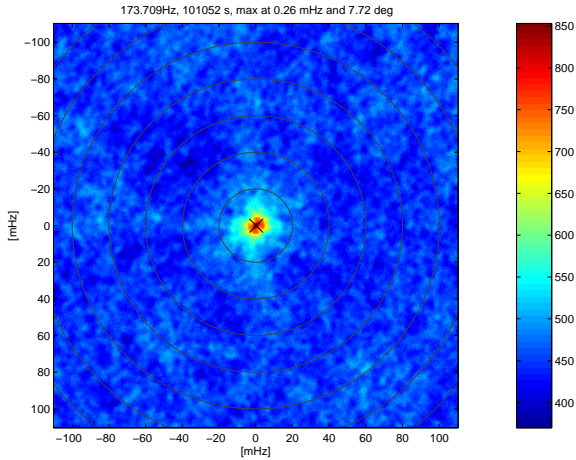,width=8.8cm,clip=} }
  \caption{47~Tucanae~C: The maximum in this Hough plane is at the
    origin, suggesting no Doppler modulation of the signal (isolated pulsar).}
  \label{fig:47TucC-hough}
\end{figure}

A closer look at the Hough planes reveals that 47~Tucanae~C at
173.709~Hz is an isolated pulsar
(Figure~\ref{fig:47TucC-hough}). Please recall that such a source
does not feature any noticeable Doppler modulation in the solar system barycenter
and the parameter $\lambda_o$ is therefore
equal to zero. This in turn is shown by the maximum at the
center/origin.

\begin{figure}
  \centering
  \centerline{\psfig{file=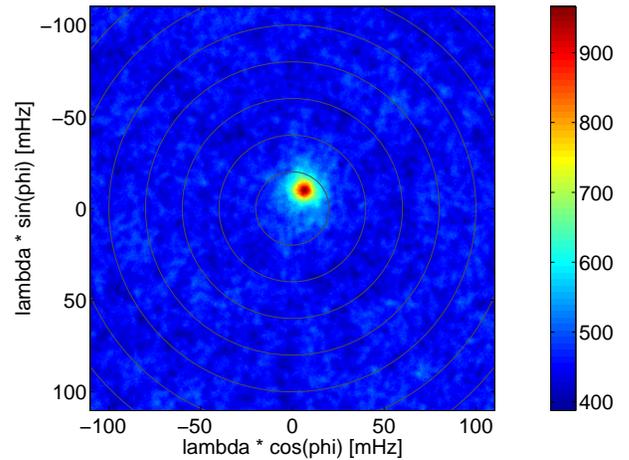,width=8.8cm,clip=} }
  \caption{47~Tucanae~J: The maximum in this Hough plane is not at the
    origin anymore. The offset indicates a Doppler modulation of
    about 12~mHz agreeing well with literature values.}
  \label{fig:47TucJ-hough}
\end{figure}

For the binary pulsar 47~Tucanae~J the situation is different. It
orbits its companion in less than 3 hours and the maximal measurable Doppler
shift for this system is about
12~mHz. Figure~\ref{fig:47TucJ-hough} shows the
Hough plane featuring the maximum at a distance to the origin
corresponding exactly to this Doppler shift.

More already known pulsars have been found in the data set, but 
are not displayed here, please refer to~\cite{Aulbert2006} for more
information.

\section*{Summary}

Although this new method has not revealed any new pulsars in the well
observer globular cluster 47~Tucanae (yet), many already known pulsars have
been ``rediscovered'' in a blind search. Since the obtained parameters
for these sources agree remarkably well with the already published
values, the method has proven to be principally able to discover new
pulsars.

The most important advantage over ``classical'' search algorithms is
the semi-coherent approach, allowing to combine data
taken on several occasions into a single, more sensitive
search. However, it should be noted that the Hough transform
requires a \textbf{large} amount of CPU power, a typical search over a
frequency bandwidth of a few hundred Hz can easily take several
10,000 CPU-hours --- the current focus is to lower this number.

Finally, the author would like to thank the pulsar group at the
Jodrell Bank observatory, especially Michael Kramer and Dunc Lorimer,
and the Albert-Einstein-Institute for the opportunity to work on this
project over the past years. At AEI especially the members of the data
analysis group\footnote{This group also uses the Hough transformation
  to search for continuous waves in data from gravitational waves detectors
  like GEO600 and LIGO. Please refer to recent papers for
  further reading, e.g.~\cite{sintes-2006-32,krishnan-2005-22,{krishnan-2004-70}}.} proved to
be invaluable sources of
knowledge. Special thanks go to Maria Alessandra Papa, Curt Cutler,
Bernard Schutz, Alicia Sintes, Steffen Grunewald, Bernd Machenschalk, Badri Krishnan, Reinhard Prix and many
more!
